\newcommand{\diracslash}[1]{#1\llap{/\kern2pt}}

\newcommand{\be}{\begin{equation}}
\newcommand{\ee}{\end{equation}}
\newcommand{\bea}{\begin{eqnarray}}
\newcommand{\eea}{\end{eqnarray}}
\newcommand{\ba}[1]{\begin{array}{#1}}
\newcommand{\ea}{\end{array}}

\newcommand{\bt}{\begin{tabular}}
\newcommand{\et}{\end{tabular}}

\newcommand{\beas}{\begin{eqnarray*}}
\newcommand{\eeas}{\end{eqnarray*}}

\documentclass[preprint,prd,aps,floats,nofootinbib,showpacs,floatfix]{revtex4}
\usepackage{graphicx}
\begin{document}

\title{D-mesons in asymmetric nuclear matter}
\author{Amruta Mishra}
\email{amruta@physics.iitd.ac.in,mishra@th.physik.uni-frankfurt.de}
\affiliation{Department of Physics, Indian Institute of Technology, Delhi,
Hauz Khas, New Delhi -- 110 016, India}

\author{Arindam Mazumdar}
\email{arindam.presi@gmail.com,arindam.mazumdar@saha.ac.in}
\affiliation{Department of Physics, Indian Institute of Technology, Delhi,
Hauz Khas, New Delhi -- 110 016, India}

\begin{abstract}
We calculate the in-medium $D$ and $\bar D$-meson masses 
in isospin aysmmetric nuclear matter 
in an effective chiral model. 
The $D$ and $\bar D$ - mass modifications
arising from  their interactions with the 
nucleons and the scalar mesons in the effective hadronic
model are seen to be appreciable at high densities
and have a strong isospin dependence.
These mass modifications can open the
channels of the decay of the charmonium states
($\Psi^\prime$, $\chi_c$, $J/\Psi$) 
to $D \bar D$ pairs in the dense hadronic matter. 
The isospin asymmetry in the doublet $D=(D^0,D^+)$
is seen to be particularly appreciable at high densities
and should show in observables like their production
and flow in asymmetric heavy ion collisions
in the compressed baryonic matter  experiments
in the future facility of FAIR, GSI.
The results of the present work are compared to
calculations of the $D(\bar D$) in-medium 
masses in the literature using the QCD sum rule 
approach, quark meson coupling model, coupled channel
approach as well as
from the studies of quarkonium dissociation using 
heavy quark potentials from lattice QCD at finite 
temperatures.

\end{abstract}
\maketitle

\def\bfm#1{\mbox{\boldmath $#1$}}

\section{Introduction}
The in-medium properties of hadrons is a topic of intense 
research in high energy physics since the last few decades.
The ongoing and future relativistic heavy ion collision experiments,
at the high energy accelerators SPS, CERN, Switzerland;
SIS, GSI, Germany; RHIC, BNL, USA; LHC, CERN, Switzerland, 
and the compressed baryonic matter (CBM) experiments at the
future facilities at GSI, Germany, 
are intended to probe matter at high temperatures and densities.
The hadrons modified in the hot and dense hadronic medium
resulting from heavy ion collision experiments, affect the
experimental observables. E.g.,the dilepton spectra
observed from heavy ion collision experiments at the SPS
\cite{ceres,helios} are attributed to the
medium modifications of vector mesons 
\cite{Brat1,CB99,vecmass,dilepton,liko},
and can not be explained by vacuum hadronic properties.
The production of kaons and antikaons as well as
their collective flow are attributed to the 
modifications of their spectral functions in the medium 
\cite{CB99,cmko,lix,Li2001,K5,K6,K4,kaosnew}.
Due to the presence of the light quark/antiquark,
the medium modifications of the $D$ ($\bar D$) mesons
can be appreciable in the  dense hadronic medium
\cite{arata,liuko,friman,weise,qcdsum08,digal,qmc}.
The observation  of open charm enhancement
in nuclear collisions \cite{cassing,lin} as well as
J/$\Psi$ suppression as observed at the SPS
\cite{NA501,NA50e,NA502}, can be due to medium
modifications of the $D$ ($\bar D$) in the medium.
In higher energy heavy ion collision experiments
at RHIC as well as at LHC, the $J/\Psi$ suppression 
can arise from formation of a quark-gluon plasma (QGP) 
\cite{satz,blaiz}.
However, the effect of the hadron absorption of $J/\Psi$
is not negligible \cite{zhang,brat5,elena}. With the drop 
in the mass of $D \bar D$ pair in the medium, the excited states 
of $J/\Psi$, a major source of yield of 
$J/\Psi$ \cite{pAdata}, can decay to the $D(\bar D)$ final states
\cite{brat6}, thereby leading to $J/\Psi$ suppression \cite{jpsi}
in the hadronic medium. The medium modification of the masses of
$J/\Psi$ and their excited states in the hadronic medium, 
due to the D-meson mass modifications have also been considered
in the literature \cite{leeko} and the excited states of $J\Psi$ 
are seen to have appreciable mass dop in the medium.
The effects due to the mass modifications of the D-mesons as well as
the charmonium bound states on experimental observables
could be explored at the future accelerator facility at GSI 
\cite{gsi}. It is thus important to study the modifications 
of the charmed mesons in the medium and hence one has to understand 
the charmed meson interactions in the hadronic phase.

In the QCD sum rule approach, due to the presence of the light quark/antiquark,
the mass modification of the $D$-mesons arises from the light quark 
condensate \cite{arata,weise,qcdsum08}. In the quark meson coupling 
(QMC) model, the contribution from the $m_c \langle \bar q q \rangle_N$ term
is represented by a quark-$\sigma$ meson coupling. The QMC model
predicts the mass shift of the $D$-meson to be of the order of 60 MeV at
nuclear matter density \cite{qmc}, which is very similar to the value
obtained in the QCD sum rule calculations of Ref. \cite{arata,weise}.
Furthermore, lattice calculations for heavy quark potentials at finite
temperature suggest a similar drop \cite{digal,lattice}.

In this work we study the medium modification of the masses of open charm
mesons (D$^\pm$) due to their interactions with the isoscalar-scalar mesons
($\sigma$ and $\zeta$ mesons), isovector-scalar $\delta$ mesons and the
nucleons in asymmetric nuclear matter. The medium modification of the 
properties of kaons and antikaons in (isospin asymmetric)
dense nuclear matter has been studied in a chiral $SU(3)$-flavor model in
Ref. \cite{isoamss,isoamss1} and have been extended to include the
effects from hyperons in the asymmetric strange hadronic matter
in ref. \cite{isoamss2}. We generalise the model to $SU(4)$-flavor to
derive the interactions of the $D$ mesons with the nucleons
and scalar mesons and investigate their mass modifications in the
asymmetric matter. The masses of the $D^{\pm}$ mesons in symmetric nuclear
matter have been studied earlier in such an effective chiral model 
\cite{amdmeson}. In a coupled channel approach  for the study of D mesons,
using a separable potential, it was shown that the resonance $\Lambda_c (2593)$
is generated dynamically in the I=0 channel \cite{ltolos} analogous to
$\Lambda (1405)$ in the coupled channel approach for the $\bar K N$ interaction
\cite{kbarn}. The approach has been generalized to study the spectral density 
of the D-mesons at finite temperatures and densitites \cite{ljhs},
taking into account the modifications of the nucleons in the medium. 
The results of this investigation seem to indicate a dominant increase 
in the width of the D-meson whereas there is only a very small change
in the D-meson mass in the medium \cite{ljhs}. However, these calculations
\cite{ltolos,ljhs}, assume the interaction to be SU(3) symmetric in u,d,c quarks
and ignore channels with charmed hadrons with strangeness.
A coupled channel approach for the study of D-mesons has been developed based
on SU(4) symmetry \cite{HL} to construct the effective interaction between 
pseudoscalar mesons in a 16-plet with baryons in 20-plet representation
through exchange of vector mesons and with KSFR condition \cite{KSFR}.
This model \cite{HL} has been modified in aspects like regularization method
and has been used to study DN interactions in Ref. \cite{mizutani6}. 
This reproduces the resonance  $\Lambda_c (2593)$ in the I=0 channel and in
addition generates another resonance in the I=1 channel at around 2770 MeV.
These calculations have been generalized to finite temperatures \cite{mizutani8} 
accounting for the in-medium modifications of the nucleons in a Walecka type
$\sigma-\omega$ model, to study the $D$ and $\bar D$ properties \cite{MK} 
in the hot and dense hadronic matter.

Within the effective chiral model considered in the present investigation, 
the $D(\bar D)$ energies are modified due to a vectorial Weinberg-Tomozawa, 
due to scalar exchange terms ($\sigma,\zeta, \delta$) as well as range terms
\cite{isoamss1,isoamss2}.
The isospin asymmetric effects among $D^0$ and $D^+$ in the doublet,
D$\equiv (D^0,D^+)$ as well as between $\bar {D^0}$ and $D^-$ in the
doublet, $\bar D \equiv (\bar {D^0},D^-)$ arise due to the scalar-isovector
$\delta$ meson, due to asymmetric contributons in the Weinberg-Tomozawa
term, as well as in the range term \cite{isoamss}.

We organize the paper as follows: We briefly recapitulate the
$SU(3)$-flavor chiral model adopted for the description of the asymmetric
hadronic matter \cite{isoamss1,isoamss2} in Section II. The properties 
then are studied within this approach. These give rise to medium modifications
for the $D$-masses through their interactions with the nucleons and scalar
mesons as presented in Section III. Section IV discusses the results of the 
present investigation, while we summarise our findings and discuss possible 
outlook in Section V.

\section{ The hadronic chiral $SU(3) \times SU(3)$ model }
The effective hadronic chiral Lagrangian used in the present work is given as
\be
{\cal L} = {\cal L}_{kin} + \sum_{ W =X,Y,V,{\cal A},u }{\cal L}_{BW}
          + {\cal L}_{vec} + {\cal L}_0 + {\cal L}_{SB}
\label{genlag} \ee are discussed. Eq. (\ref{genlag}) corresponds
to a relativistic quantum field theoretical model of baryons and
mesons adopting a nonlinear realization of chiral symmetry
\cite{weinberg,coleman,bardeen} and broken scale invariance (for
details see \cite{paper3,hartree,kristof1}) as a description of 
the hadronic matter. The model was used successfully to
describe nuclear matter, finite nuclei, hypernuclei and neutron
stars. The Lagrangian contains the baryon octet, the spin-0 and
spin-1 meson multiplets as the elementary degrees of freedom. In
Eq. (\ref{genlag}), $ {\cal L}_{kin} $ is the kinetic energy term,
$  {\cal L}_{BW}  $ contains the baryon-meson interactions in
which the baryon-spin-0 meson interaction terms generate the
baryon masses. $ {\cal L}_{vec} $ describes the dynamical mass
generation of the vector mesons via couplings to the scalar fields
and contains additionally quartic self-interactions of the vector
fields.  ${\cal L}_0 $ contains the meson-meson interaction terms
inducing the spontaneous breaking of chiral symmetry as well as a
scale invariance breaking logarithmic potential. $ {\cal L}_{SB} $
describes the explicit chiral symmetry breaking.

The baryon-scalar meson interactions generate the baryon masses
and the parameters corresponding to these interactions are adjusted 
so as to obtain the baryon masses as their experimentally measured 
vacuum values. For the baryon-vector meson interaction terms, there
exist the $F$-type (antisymmetric) and $D$-type (symmetric) couplings.
Here we will use the antisymmetric coupling \cite{paper3,isoamss1,isoamss2}
because, following the universality principle  \cite{saku69}
and the vector meson dominance model, one can conclude that
the symmetric coupling should be small.
Additionally we choose the parameters \cite{paper3,isoamss} so as to
decouple the strange vector field $
\phi_\mu\sim\bar{s} \gamma_\mu s $ from the nucleon,
corresponding to an ideal mixing between $\omega$ and $\phi$.
A small deviation of the mixing angle from the ideal mixing
\cite {dumbrajs,rijken,hohler1} has not been taken into
account in the present investigation.

The Lagrangian densities corresponding to the interaction
for the vector meson, ${\cal L}_{vec}$, the meson-meson interaction
${\cal L}_0$ and that corresponding to the explicit chiral symmetry breaking
${\cal L}_{SB}$ have been described in detail in references
\cite{paper3,isoamss}.

To investigate the hadronic properties in the medium, we write
the Lagrangian density within the chiral SU(3) model in the mean
field approximation and determine the expectation values
of the meson fields by minimizing the thermodynamical potential
\cite{hartree,kristof1}.

\section{$D$ and $\bar D$ mesons in the medium}

We now examine the medium modifications for the $D$ and $\bar D$-meson 
masses in the asymmetric nuclear matter. The properties of nucleons 
and scalar mesons have been studied in the asymmetric hadronic matter
within a chiral SU(3) model \cite{isoamss1}.
We assume that the additional effect of charmed particles in the medium
leads to only marginal modifications \cite{roeder} of these hadronic
properties and do not need to be taken into account here.
However, to investigate the medium modification of the $D$-meson mass,
we need to know the interactions of the $D$-mesons with the light
hadron sector.

The light quark condensate has been shown to
play an important role for the shift in the $D$-meson mass in
the QCD sum rule calculations \cite{arata}. In the present chiral model,
the interactions to the scalar fields (nonstrange, $\sigma$ and strange,
$\zeta$) as well as a vectorial Weinberg-Tomozawa interaction
term modify the masses for D$^\pm$ mesons in the medium.
These interactions were considered within the SU(3) chiral model
to investigate the modifications of K-mesons in the dense (asymmetric)
hadronic medium \cite{isoamss,isoamss1,isoamss2}. 

To consider the medium effects on the $D$ and $\bar D$-meson masses
we generalize the chiral $SU(3)$-flavor model to include the charmed 
mesons. The scalar meson multiplet has now the expectation value
\begin{equation}
\langle X \rangle 
= \left(
\begin{array}{cccc} (\sigma+\delta)/\sqrt 2 & 0 & 0 & 0\\
 0 & (\sigma-\delta)/\sqrt 2 & 0 & 0 \\
 0 &  0 & \zeta & 0 \\
 0 &  0 & 0 & \zeta_c \\
\end{array}\right),
\end{equation}
with $\zeta_c$ corresponding to the $\bar c c$--condensate.
The pseudoscalar meson field P can be written, including the charmed
mesons, as
\begin{equation}
P = \left(
\begin{array}{cccc} \frac {\pi^0}{\sqrt 2} & \pi^+ & \frac{2 K^+}{1+w} & 
\frac {2 {\bar D}^0}{1+w_c} \\
\pi^- & -\frac {\pi^0}{\sqrt 2} & 
\frac{2 K^0}{1+w} & 
\frac {2 D^-}{1+w_c} \\
\frac {2 K^-}{1+w} & 
\frac {2 {\bar K^0}}{1+w} & 
0 & 0 \\
\frac {2 D^0}{1+w_c}
& \frac {2 D^+}{1+w_c} & 0 & 0 \end{array}\right),
\end{equation}
where $w=\sqrt 2 \zeta/\sigma$ and  $w_c=\sqrt 2 \zeta_c/\sigma$.
From PCAC, one gets the decay constants for the pseudoscalar mesons
as $f_\pi=-\sigma$, $f_K=-(\sigma +\sqrt 2 \zeta )/2$ and
$f_D=-(\sigma +\sqrt 2 \zeta_c )/2$. In the present calculations,
the value for the D-decay constant will be taken to be 135 MeV
\cite{weise}. We note that for the decay constant of $D_s^+$,
the Particle Data Group \cite{pdg} quotes a value of 
$f_{D_s^+} \simeq 200 MeV$. Taking a similar value also for
$f_D$ would not affect our results qualitatively, however
(see also \cite{lat03}).

The interaction Lagrangian modifying the $D$-meson mass can be written
as \cite{isoamss1}
\begin{eqnarray}
\cal L _{DN} & = & -\frac {i}{8 f_D^2} \Big [3\Big (\bar p \gamma^\mu p
+\bar n \gamma ^\mu n \Big) 
\Big({D^0} (\partial_\mu \bar D^0) - (\partial_\mu {{D^0}}) {\bar D}^0 \Big )
+\Big(D^+ (\partial_\mu D^-) - (\partial_\mu {D^+})  D^- \Big )
\nonumber \\
& +&
\Big (\bar p \gamma^\mu p -\bar n \gamma ^\mu n \Big) 
\Big({D^0} (\partial_\mu \bar D^0) - (\partial_\mu {{D^0}}) {\bar D}^0 \Big )
- \Big( D^+ (\partial_\mu D^-) - (\partial_\mu {D^+})  D^- \Big )
\Big ]
\nonumber \\
 &+ & \frac{m_D^2}{2f_D} \Big [ 
(\sigma +\sqrt 2 \zeta_c)\big (\bar D^0 { D^0}+(D^- D^+) \big )
 +\delta \big (\bar D^0 { D^0})-(D^- D^+) \big )
\Big ] \nonumber \\
& - & \frac {1}{f_D}\Big [ 
(\sigma +\sqrt 2 \zeta_c )
\Big ((\partial _\mu {{\bar D}^0})(\partial ^\mu {D^0})
+(\partial _\mu {D^-})(\partial ^\mu {D^+}) \Big )
\nonumber \\
 & + & \delta
\Big ((\partial _\mu {{\bar D}^0})(\partial ^\mu {D^0})
-(\partial _\mu {D^-})(\partial ^\mu {D^+}) \Big )
\Big ]
\nonumber \\
&+ & \frac {d_1}{2 f_D^2}(\bar p p +\bar n n 
 )\big ( (\partial _\mu {D^-})(\partial ^\mu {D^+})
+(\partial _\mu {{\bar D}^0})(\partial ^\mu {D^0})
\big )
\nonumber \\
&+& \frac {d_2}{4 f_D^2} \Big [
(\bar p p+\bar n n))\big ( 
(\partial_\mu {\bar D}^0)(\partial^\mu {D^0})
+ (\partial_\mu D^-)(\partial^\mu D^+) \big )\nonumber \\
 &+&  (\bar p p -\bar n n) \big ( 
(\partial_\mu {\bar D}^0)(\partial^\mu {D^0})\big )
- (\partial_\mu D^-)(\partial^\mu D^+) ) 
\Big ]
\label{lagd}
\end{eqnarray}
In (\ref{lagd}) the first term is the vectorial interaction term
obtained from the kinetic term in (\ref{genlag}). The second term, 
which gives an attractive interaction for the $D$-mesons, is obtained 
from the explicit symmetry breaking term in (\ref{genlag}).
The third term arises from kinetic term of the pesudoscalar mesons
\cite{isoamss1,isoamss2}. The fourth and fifth terms have been written down 
for the DN interactions, in analogy with the $d_1$ and $d_2$ terms 
in chiral SU(3) model \cite{isoamss1,isoamss2}.
The last three terms in (\ref{lagd})
represent the range term in the chiral model. 
It might be noted here that the interaction of the pseudoscalar
mesons to the vector mesons, in addition to the pseudoscalar meson--nucleon
vectorial interaction leads to a double counting in the linear realization
of the chiral effective theory \cite{borasoy}. Within the
nonlinear realization of the chiral effective theories, such an interaction
does not arise in the leading or sub-leading order, but only as a higher
order contribution \cite{borasoy}. Hence the vector meson-pesudoscalar
interaction will not be considered within the present investigation.

The Fourier transformations of the equations of motion for 
$D$ and $\bar D$ mesons yield the dispersion relations,
\begin {equation}
\omega^2+ {\vec k}^2 + m_D^2 -\Pi(\omega, |\vec k|)=0,
\label {dispdm}
\end{equation}
where $\Pi$ denotes the self energy of the $D$ ($\bar D$) meson
in the medium.

Explicitly, the self energy $\Pi (\omega,|\vec k|)$ for the $D$ meson doublet,
($D^0$,$D^+$) 
arising from the interaction (\ref{lagd}) is given as
\begin{eqnarray}
\Pi (\omega, |\vec k|) &= & \frac {1}{4 f_D^2}\Big [3 (\rho_p +\rho_n)
\pm (\rho_p -\rho_n) \big)
\Big ] \omega \nonumber \\
&+&\frac {m_D^2}{2 f_D} (\sigma ' +\sqrt 2 {\zeta_c} ' \pm \delta ')
\nonumber \\ & +& \Big [- \frac {1}{f_D}
(\sigma ' +\sqrt 2 {\zeta_c} ' \pm \delta ')
+\frac {d_1}{2 f_D ^2} (\rho_s ^p +\rho_s ^n)\nonumber \\
&+&\frac {d_2}{4 f_D ^2} \Big (({\rho^s} _p +{\rho^s} _n)
\pm   ({\rho^s} _p -{\rho^s} _n) \Big ) \Big ]
(\omega ^2 - {\vec k}^2),
\label{selfdm}
\end{eqnarray}
where the $\pm$ signs refer to the $D^0$ and $D^+$ respectively.
In the above, $\sigma'(=\sigma-\sigma _0)$,
${\zeta_c}'(=\zeta_c-{\zeta_c}_0)$ and  $\delta'(=\delta-\delta_0)$
are the fluctuations of the scalar-isoscalar fields $\sigma$ and $\zeta_c$,
and the third component of the scalar-isovector field, $\delta$,
from their vacuum expectation values.
The vacuum expectation value of $\delta$ is zero ($\delta_0$=0), since
a nonzero value for it will break the isospin symmetry of the vacuum
( we neglect here the small isospin breaking effect arising from 
the mass and charge difference of the up and down quarks).
In the above, $\rho_p$ and $\rho_n$ are the number densities
of proton and neutron and ${\rho^s}_{p}$ and ${\rho^s}_n$ are
their scalar densities.

Similarly, for the $\bar D$ meson doublet, (${\bar D}^0$,$D^-$),
the self-energy is calculated as
\begin{eqnarray}
\Pi (\omega, |\vec k|) &= & -\frac {1}{4 f_D^2}\Big [3 (\rho_p +\rho_n)
\pm (\rho_p -\rho_n) \Big ] \omega\nonumber \\
&+&\frac {m_D^2}{2 f_D} (\sigma ' +\sqrt 2 {\zeta_c} ' \pm \delta ')
\nonumber \\ & +& \Big [- \frac {1}{f_D}
(\sigma ' +\sqrt 2 {\zeta_c} ' \pm \delta ')
+\frac {d_1}{2 f_D ^2} (\rho_s ^p +\rho_s ^n
)\nonumber \\
&+&\frac {d_2}{4 f_D ^2} \Big (({\rho^s} _p +{\rho^s} _n)
\pm   ({\rho^s} _p -{\rho^s} _n) \Big ]
(\omega ^2 - {\vec k}^2),
\label{selfdbarm}
\end{eqnarray}
where the $\pm$ signs refer to the $\bar {D^0}$ and $D^-$ respectively.

The optical potentials are calculated from the energies of the $D$ 
and $\bar D$ mesons
\be
U(\omega, k) = \omega (k) -\sqrt {k^2 + m_D ^2},
\ee
where $m_D$ is the vacuum mass for the $D(\bar D)$ meson
and $\omega (k)$ is the momentum dependent energy of
the $D (\bar D)$ meson.

The parameters $d_1$ and $d_2$ are determined by a fit of the
empirical values of the KN scattering lengths
\cite{thorsson,juergen,barnes}
for I=0 and I=1 channels \cite{isoamss1,isoamss2}.

In the next section, we shall discuss the results for 
the $D$-meson mass modification obtained in the present effective 
chiral model as compared to the results in the literature,
obtained from other approaches.

\section{Results and Discussions}
\label{results}
To study the $D(\bar D)$-meson masses in  asymmetric nuclear medium
due to its interactions with the light hadrons,
we have generalized the chiral SU(3) model
used for the study of the dense hadronic matter to SU(4)
for the meson sector.
The contributions from the various terms of the interaction Lagrangian
(\ref{lagd}) to the masses of the $D\equiv (D^0,D^+)$ and
$\bar D \equiv ({\bar {D^0}},D^-$) are shown in figures 1 and 2,
as functions of density. These are illustrated for the isospin asymmetric case
with the value of the asymmetry parameter, $\eta=(\rho_n-\rho_p)/(2 \rho_B)$ 
as 0.5 and compared with the results obtained for the isospin
symmetric case ($\eta$=0).
The isospin symmetric part ($\sim (\rho_n +\rho_p$)) of the first term 
of (\ref{lagd}), called the Weinberg-Tomozawa
term, is attractive for $D \equiv (D^0,D^+)$ mesons and leads 
to a drop of the masses
of the $D^+$ and $D^0$ mesons, whereas it is repulsive for
the $\bar D$ mesons in the nuclear medium, leads to an increase
of the $\bar D^0$ and $D^-$ meson masses. In the isospin asymmetric 
nuclear medium,
the Weinberg-Tomozawa term, leads to a mass splitting of the $D^0$ 
and $D^+$ mesons, giving a further drop in the mass of $D^+$,
whereas the asymmetry reduces the drop of the mass in $D^0$.
The D-meson self energy arising from the Weinberg-Tomozawa 
interaction, $\Pi_{WT} (\omega,|\vec k|)$ is given by the first 
term of equation (\ref{selfdm}), and at low densities, this turns 
out to be much smaller than $(\vec k ^2 +m_D ^2)$. One can then, 
as a first approximation replace $\Pi _{WT} (\omega,|\vec k|)$ 
by $\Pi_{WT} (m_D,|\vec k|)$ 
and solve for dispersion relation given by (\ref{dispdm}).
Confining our attention to the Weinberg-Tomozawa interaction only, 
the energies of the $D^0$ and $D^+$ mesons,
in the above approximation, are given by
\begin{equation}
\omega (|\vec k|)\simeq (|\vec k|^2+m_D^2)^{1/2}
-\frac {1}{8 f_D^2} \Big [3(\rho_p +\rho_n)\mp (\rho_p -\rho_n)\Big ]
\label{endman}
\end{equation}
One can note from the above equation (\ref{endman}) that 
at low densities, a given isospin asymmetry introduces equal increase 
(drop) for mass of $D^0$($D^+$) meson at low densities. 
However, at higher densities,
there are deviations from the analytical expressions given 
by (\ref{endman}) as expected, though the qualitative feature
of the $D^+$ ($D^0$) experiencing an attractive (repulsive)
interaction from the vectorial Weinberg-Tomozawa term of 
(\ref{selfdm}) still remains the same, as can be seen from
figure 1. One sees, from figure 2, that there is an increase 
in the masses of  the $D^-$ and $\bar {D^0}$ and their mass shifts 
are equal for the case of isospin symmetric matter ($\eta$=0).
However, in the presence of asymmetry, the $D^-$ mass is 
seen to have an increase whereas $\bar {D^0}$ mass drops
in the asymmetric medium. These behaviours for $D^-$ and $\bar {D^0}$
can be understood by examining the asymmetric contributions of the 
Weinberg-Tomozawa term (the first term of (\ref{selfdbarm})).

The scalar meson exchange contribution to the self energy
is given by the second term of equation (\ref{selfdm}) for $D$ mesons 
and by the second term of (\ref{selfdbarm}) for $\bar D$ mesons. 
Its interaction is attractive and is identical for both the $D$ 
and $\bar D$ doublets for the isospin symmetric nuclear matter
(a negligible difference in the energies is due to the difference
in their vacuum masses, $m_{D^+}=m_{D^-}$=1869 MeV and
$m_{D^0}=m_{\bar {D^0}}$=1864.5 MeV). One might notice 
from the self energy terms due to the scalar meson exchange that
a nonzero value of the scalar isovector $\delta$-meson arising 
due to isospin asymmetry in nuclear matter, gives a drop in the 
masses for the $D^+$ and $D^-$, whereas the $\delta$ contribution 
is repulsive for $D^0$ and $\bar {D^0}$ ($\sigma'=\sigma-\sigma_0 >0$
and $\delta '=\delta <0$). 
This gives the $D^0$ mass to be higher than $D^+$ mass for 
asymmetric nuclear matter as seen in figure 1 and the mass
of $\bar {D^0}$ mass (identical to $\bar D^0$) to be higher than 
$D^-$ mass (identical to $D^+$ mass) as plotted in figure 2. 
However, one might notice that the shifts 
in the $D^+$($D^-$) and $D^0$($\bar {D^0}$)  masses 
about the isospin symmetric case ($\eta$=0) case are not 
equal and opposite. The reason for the seen asymmetry 
in the mass splittings is the following. For low baryon 
densities, one has $\sigma' \sim \rho_s \simeq \rho_B$,
and $\delta' \sim (\rho_p^s-\rho_n^s) \simeq (\rho_p-\rho_n)$,
so that one would expect the splittings of the masses of
$D^+$ and $D^0$ to be symmetrical (about the isospin symmetric matter)
for a given baryon density $\rho_B$ and isospin asymmetry parameter, $\eta$.
However, these no longer hold good for higher densities
and the mean field $\sigma$ calculated in the isospin symmetric 
situation ($\eta$=0) turns out to be different
for the asymmetric situation when the coupled equations of motions
are solved for the scalar mean fields due to the presence of
the scalar isovector $\delta$ field, as compared to when these
equations are solved for symmetric nuclear matter (for $\eta$=0)  
in the absence of the $\delta$ meson. 

The contributions to the $D$ and $\bar D$ self energies due to
range terms are given by the last three terms of the right hand side
of equations (\ref{selfdm}) and (\ref{selfdbarm}) respectively.
The first term of the range terms given by the third term in (\ref{selfdm})
is repulsive whereas the second and third range terms have
attractive contributions, when the isospin asymmetry is not taken into
account. However, the isospin asymmetry, due to a nonzero value of
the $\delta$ field, leads to an increase in the masses of the $D^+$ 
and $D^-$ mesons and a drop in the masses of $D^0$ and $\bar {D^0}$ 
from the isospin symmetric case.
The second of the range terms (the $d_1$ term) is 
attractive and gives identical mass drops for $D^+$ and $D^0$ in the
$D$ doublet as well as for $D^-$ and $\bar D^0$ in the $\bar D$ doublet.
This term is proportional to $(\rho_s^p+\rho_s^n)$, which turns
out to be different for the isospin asymmetric case 
as compared to the isospin symmetric nuclear matter, 
due to the presence of the $\delta$ meson. 
This is because the equations of motion for the scalar fields
for the two situations (with/without $\delta$ mesons) give different values
for the mean field, $\sigma$ ($\sim ({\rho_s}^p +{\rho_s}^n)$).
The last term of the range term (the $d_2$ term) has a negative
contribution for the energies of $D^+$ and $D^0$ mesons 
as well as for $D^-$ and $\bar {D^0}$ mesons for the isospin
symmetric matter. The isospin asymmetric part arising from the
($({\rho_s}^n -{\rho_s} ^p)$) term of the $d_2$ term has a further 
drop in the masses for $D^\pm$ mesons,
whereas it increases the masses of the $D^0$ and $\bar{D^0}$ mesons
from their isospin symmetric values. 
One sees from figures 1 and 2 that for both $D^+$ and $D^-$ mesons,
the modification in the masses arising from the range terms, 
for the asymmetric matter ($\eta$=0.5),
as compared to the isospin symmetric matter is negligible 
due to the increase from the first two range terms almost cancelling
with the drop due to the $d_2$ term. For the $D^0$ and $\bar D^0$ mesons,
the mass is seen to increase as compared to the isospin symmetric matter,
due to the increase due to the second and third range terms 
dominating over the drop due to first range term. 
The values of the paramters $d_1$ and $d_2$ are fitted from the
kaon-nucleon scattering lengths \cite{isoamss,isoamss1} to be
2.56/$m_K$ and 0.73/$m_K$ respectively \cite{isoamss2} and it is seen 
that the $d_2$ term has a smaller contribution as compared to 
the $d_1$ term. At high densities, these attractive $d_1$ and
$d_2$ terms dominate over the first range term (repulsive) and 
this leads to a decrease of the masses of the $D$ and $\bar D$ mesons.
There is seen to be a substantial drop of D-meson masses at high
densities due to the inclusion of this range term.

The density dependence of the  masses of the $D$ mesons at specific
values of the isospin asymmetric parameter, $\eta$, are shown
in figure 3. The isospin asymmetry is seen to give a rise (drop) 
in the masses of the $D^0$ ($D^+$) as compared to their masses
in the symmetric matter. For the isospin symmetric nuclear matter 
($\eta$=0), the drop in the mass of the $D^+$ at $\rho_B=\rho_0$ 
is about 81 MeV from its vacuum value of 1869 MeV and
$D^-$ meson mass also has a drop of about 30 MeV, giving
a mass splitting between the $D^+$ and $D^-$ mesons
as about 51 MeV. A similar drop of the D-meson
mass is also predicted by the QMC model \cite{qmc}
as well as for the isospin averaged $D(\bar D$) mass
in the QCD sum rule approach in \cite{arata}.
In the recent QCD sum rule calculations \cite{qcdsum08},
the mass drop for $D^+D^-$ is about 50 MeV and the mass splitting
is about 90 MeV at $\rho=\rho_0$, whereas for the isospin symmetric 
nuclear matter, we obtain these values as 110 MeV and 50 MeV respectively.
The coupled channel calculations \cite{ljhs} show a dominant 
modification of the D-meson width and only a small change in the
$D$ meson mass in the medium. When the DN interaction is taken to be
a Tomozawa Weinberg interaction supplemented by a scalar-isoscalar interaction
\cite{mizutani8}, the mass modification of the $D$ mesons is seen
to be around 10 MeV at $\rho_0$ and about 50 MeV for a density of $2 \rho_0$.
For the $D^-$ meson, there is an increase in the mass by about 10 MeV and
30 MeV at densities of $\rho_0$ and $2 \rho_0$ respectively.  

In the present investigation, the isospin dependence is seen
to be quite prominent for $D^0$ as compared to $D^+$.
The isospin asymmetry in the medium is seen to give an
increase in the $D^0$ mass, whereas it gives a drop
for the $D^+$ mass as compared to the $\eta$=0 case.
For the isospin asymmetry parameter, $\eta$=0.5,
the $D^0$ mass is seen to rise by about 23 MeV and 105 MeV
for $\rho_B$ as $\rho_0$ and 5$\rho_0$ respectively
from their values of 1783 MeV and 1441 MeV in isospin symmetric  
matter. For the same value of asymmetry parameter, $\eta$,
the $D^+$ is seen to have a mass drop of about 18 MeV 
and 39 MeV for $\rho_B=\rho_0$ and 5$\rho_0$ respectively, from its
$\eta$=0 values.  This strong isospin dependence of the $D$ mesons 
should show up in observables like their production 
as well as flow in asymmetric heavy ion collisions 
planned at the future facility at FAIR, GSI.
Figure 4 shows the masses of the $D^-$ and $\bar {D^0}$.
It is seen that at high densities, both the $D^-$ and $\bar {D^0}$
are seen to have an increase in their masses in the asymmetric 
nuclear matter, as compared to the isospin symmetric case
and this rise in the masses are seen to be similar for
both $D^-$ and ${\bar {D^0}}$.
For example, at nuclear matter density, the masses
of $D^-$ and ${\bar {D^0}}$ are 1838 MeV and 1840 MeV 
respectively for $\eta$=0.5, which are very similar to the
values of 1839 MeV and 1834.5 MeV for the isospin symmetric 
matter. For $\rho_B=5 \rho_0$, the masses of $D^-$ and ${\bar {D^0}}$
are 1690 MeV and 1698 MeV respectively, which are higher 
by about 31 MeV and 43 MeV from the isospin symmetric case.
One sees that the values of $D^-$ and ${\bar {D^0}}$ masses remain 
very similar at a given isospin asymmetry. But it is seen that
the density effects on these masses are quite appreciable 
(a drop of about 30 MeV for nuclear matter density and 
of about 180 MeV for a density about five times nuclear matter density
for $\eta$=0.5).

Figures 5 and 6 show the optical potentials for the
$D$ and $\bar D$ doublets as functions of the momentum.
These are shown for densities $\rho_0$ and 5$\rho_0$.
The isospin dependence of the optical potentials
is seen to be quite significant for high densities
for the doublet ($D^+,D^0$), as has also been
seen in the case of their masses. But, we see the optical
potentials for the $\bar D$ have very similar values for
$D^-$ and ${\bar {D^0}}$ for a fixed value of the isospin
asymmetric parameter. But, as already has been seen for the
case of their masses, we see the optical potentials
for both $D^-$ and ${\bar {D^0}}$ are quite different
from the symmetric nuclear matter case at high densities.

The decay widths of the charmonium states can be modified by the level
crossings between the excited states of J/$\Psi$ (i.e.,
$\Psi^\prime$,$\chi_c$) and the threshold for $D\bar D$ creation due to
the medium modifications of the $D$-meson masses \cite{friman}. In the
vacuum, the resonances above the $D\bar D$ threshold, for example the
$\Psi ''$ state, has a width of 25 MeV due to the strong open charm
channel. On the other hand, the resonances below the threshold  have a
narrow width of a few hundreds of KeV, only. With the medium
modification of the $D(\bar D)$-meson masses, the channels for the excited
states of $J/\Psi$, like $\chi_c$, $\Psi^\prime$ decaying to $D^+D^-$
or $D^0 \bar {D^0}$ pairs can open up in the dense hadronic medium.
This can increase the decay widths of $\Psi '$ and $\chi_c$ states
at high densities. In figure 7, we show the density dependence
of the mass of the $D^+ D^-$ as well as $D^0 \bar {D^0}$ pair,
calculated in the present investigation. We see that the decay to
$D^+ D^-$ channel is almost insensitive to isospin asymmetry.
On the other hand, the decay channel to the $D^0 \bar {D^0}$
is seen to have a strong isospin asymmetry dependence, with
the asymmetry shifting the onset of the decay to higher densities.
The decay of charmonium states
to $D\bar D$ has been studied in Ref. \cite{brat6,friman}. It is seen
to depend sensitively on the relative momentum in the final state.
These excited states might  become narrow \cite{friman} though the
$D$-meson mass is decreased appreciably at high densities. 
It may even vanish at certain momenta corresponding to
nodes in the wavefunction \cite{friman}. Though the decay widths for
these excited states can be modified by their wave functions, the
partial decay width of $\chi_{c2}$, due to absence of any nodes, can
increase monotonically with the drop of the $D^+ D^-$ pair mass in the
medium \cite{friman}. This can give rise to depletion in the $J/\Psi$
yield in heavy ion collisions. The dissociation of the quarkonium
states ($\Psi^\prime$, $\chi_c$, $J/\Psi$) into $D\bar D$ pairs have 
also been studied \cite{digal,wong} by comparing their binding energies 
with lattice results on temperature dependence of heavy quark effective
potential \cite{lattice}. 

\section{summary}
To summarize we have investigated in a chiral model the in-medium
masses of the $D, \bar D$-mesons in asymmetric nuclear matter,
arising due to their interactions with the nucleons and scalar mesons. 
The properties of the light hadrons -- as studied in a $SU(3)$ chiral model
-- modify the $D (\bar D)$-meson properties in the dense hadronic
medium. The SU(3) model with paramaters fixed from the properties of
hadron masses in the vacuum, and low energy  KN scattering data,
is extended to SU(4) to derive the interactions of $D (\bar D)$-mesons 
with the light hadron sector. The mass modifications for the $D$ mesons 
are seen to be similar to earlier finite density calculations
of QCD sum rules \cite{weise,qcdsum08} as well as to the quark meson coupling
(QMC) model \cite{qmc}, in contrast to the small mass modifications in the
coupled channel approach \cite{ljhs,mizutani8}. In our calculations, 
the presence of the repulsive range term (the fourth term
of (\ref{lagd}) is compensated by the attractive 
$d_1$ and $d_2$ terms given by the last two terms in (\ref{lagd}) 
and the latter (dominated by the $d_1$ term) have an effect of reducing the
masses of both $D$ and $\bar D$ mesons at high densities. 

The medium modification of the $D$-masses can lead to a suppression in the
$J/\Psi$- yield in heavy-ion collisions, since the excited states
of $J/\Psi$ and at a much higher density ($\simeq 5\rho_0$),
$J/\Psi$ can decay to $D\bar D$ pairs in the dense hadronic medium.
The decay to $D^+ D^-$ pair seems to be insensitive to isospin
dependence, whereas due to increase in the mass of $D^0 \bar {D^0}$
in the asymmetric medium, isospin asymmetry is seen to disfavour
the decay of the charmonium states to $D^0 \bar {D^0}$ pair.
The isospin dependence of $D^+$ and $D^0$ masses is seen to be 
dominant medium effect at high densities which might show in
their production ($D^+/D^0$), whereas for the $D^-$ and $\bar {D^0}$,
one does see that even though these have a strong density dependence,
their in-medium masses remain similar at a given value for
the isospin asymmetry parameter $\eta$.
The strong density dependence as well as the isospin dependence
of the D($\bar D$) meson optical potentials in the asymmetric nuclear
matter can be tested in the asymmetric heavy ion collision experiments
at the future GSI facility \cite{gsi}.

\acknowledgements
We thank Arvind Kumar, Sambuddha Sanyal, Laura Tolos and J. Schaffner-Bielich
for fruitful discussions. 
One of the authors (Amruta Mishra) is grateful to the Institut 
f\"ur Theoretische Physik for warm hospitality and acknowledges financial 
support from Alexander von Humboldt Stiftung when this work was initiated. 
Financial support from Department of Science and Technology, Government 
of India (project no. SR/S2/HEP-21/2006) is also gratefully acknowledged.

\begin{figure}
\includegraphics[width=16cm,height=16cm]{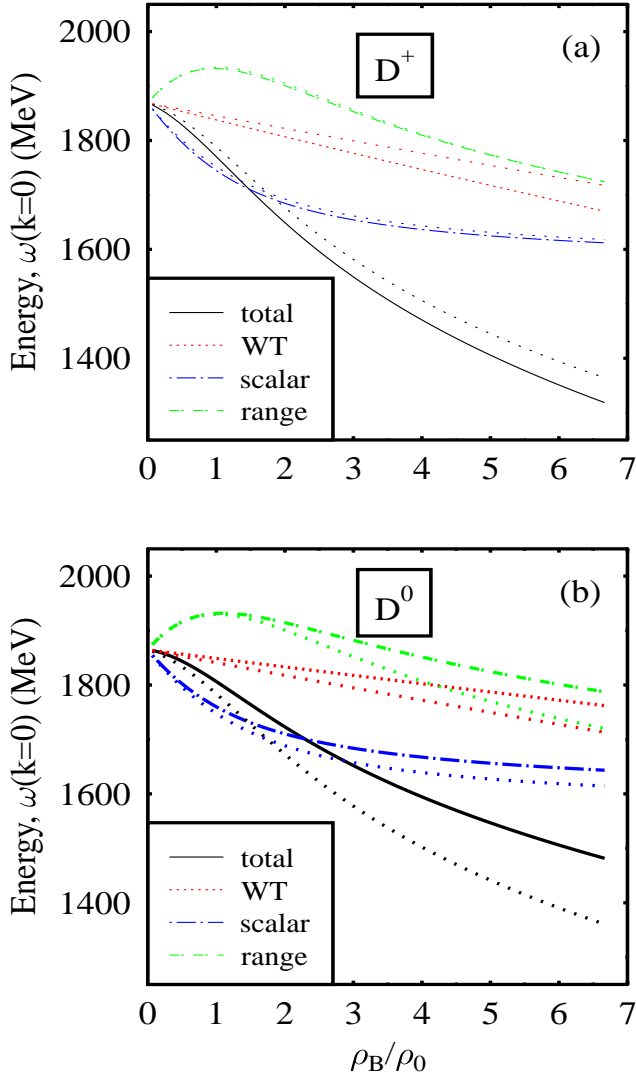}
\caption{(Color online)
The various contributions to D meson energies at zero momentum 
(for $D^+$ in (a) and for $D^0$ in (b)) in MeV plotted
as functions of the baryon density in units of nuclear matter
saturation density, $\rho_B/\rho_0$
are shown for the isospin asymmetry  parameter, $\eta$=0.5 
and compared with the case of $\eta$=0 (dotted line).
}
\end{figure}

\begin{figure}
\includegraphics[width=16cm,height=16cm]{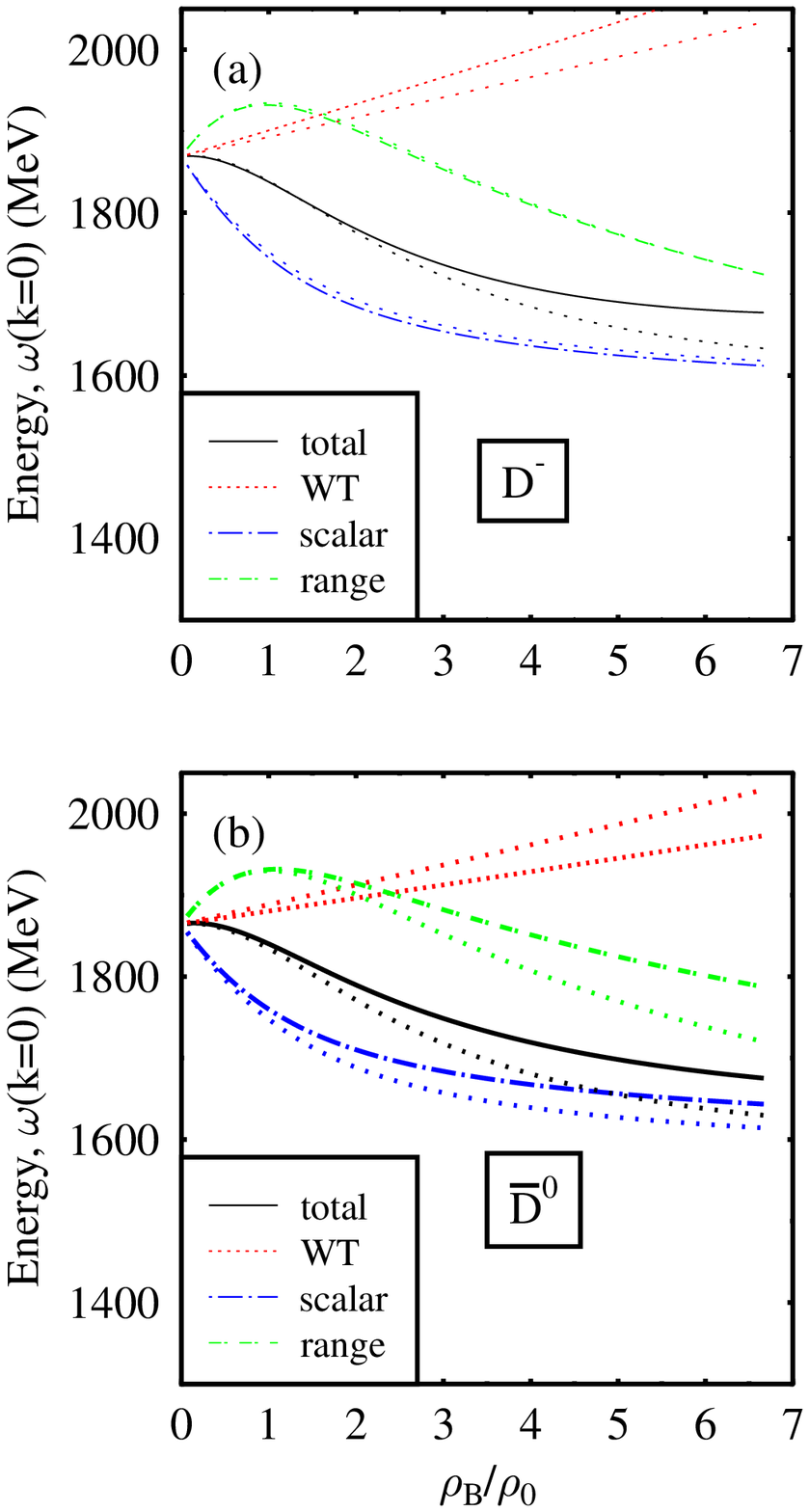}
\caption{(Color online)
The variuos contributions to $\bar D$ meson energies at zero momentum
(for $D^-$ in (a) and for $\bar {D^0}$ in (b)) in MeV plotted
as functions of the baryon density in units of nuclear matter
saturation density, $\rho_B/\rho_0$
are shown for the isospin asymmetry  parameter, $\eta$=0.5 
and compared with the case of $\eta$=0 (dotted line).
}
\end{figure}

\begin{figure}
\includegraphics[width=16cm,height=16cm]{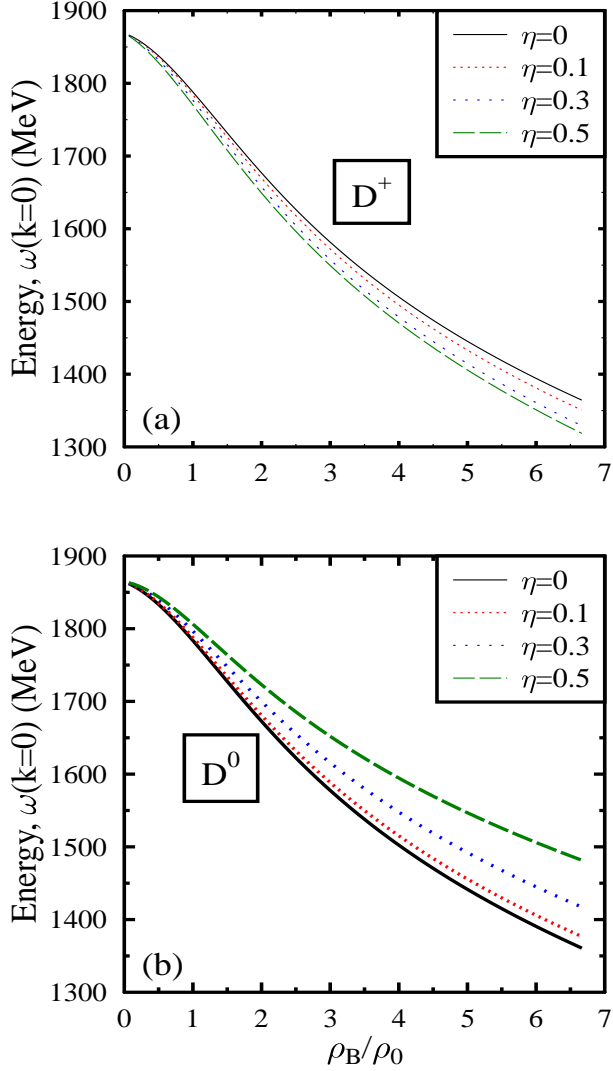}
\caption{(Color online)
The D meson energies for zero momentum (k=0) (for $D^+$ in (a) 
and for $D^0$ in (b)) 
in MeV plotted as functions of the baryon density in units
of $\rho_0$, $\rho_B/\rho_0$
for different values of the isospin asymmetry  parameter, $\eta$.
}
\end{figure}

\begin{figure}
\includegraphics[width=16cm,height=16cm]{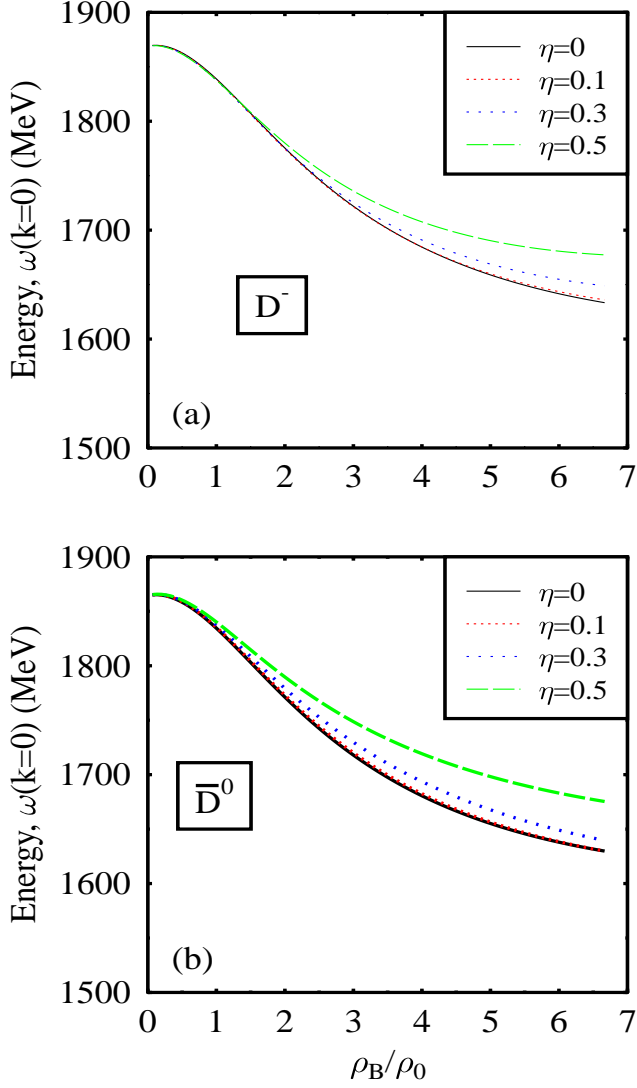}
\caption{(Color online)
The energies of the $\bar D$ mesons for zero momentum 
(for $D^-$ in (a) and for $\bar {D^0}$ 
in (b)) as functions of the baryon density in units of $\rho_0$
($\rho_B/\rho_0$), are plotted for different values of the isospin 
asymmetry parameter, $\eta$.
}
\end{figure}

\begin{figure}
\includegraphics[width=16cm,height=16cm]{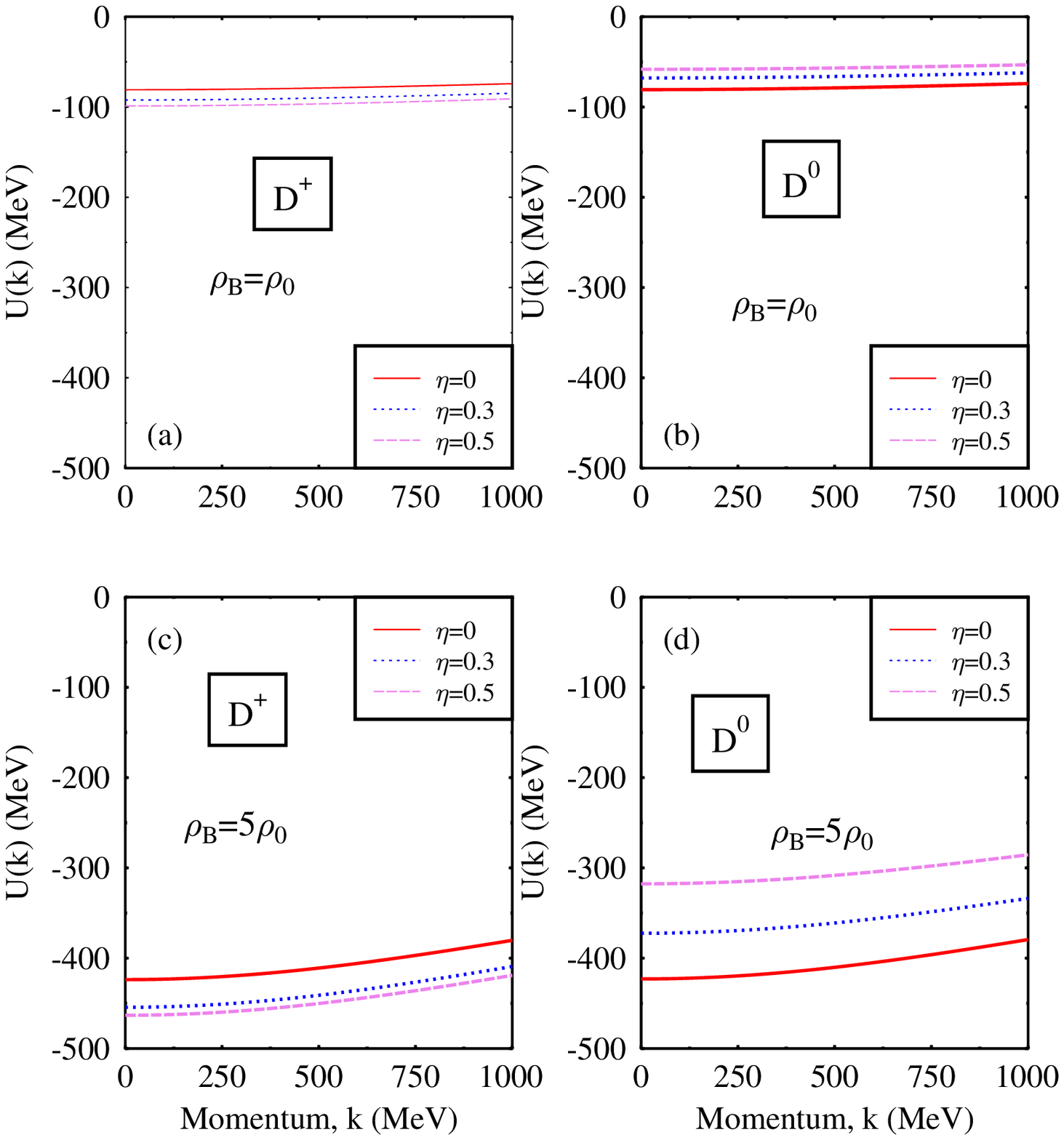}
\caption{(Color online)
The optical potentials (for $D^+$ and $D^0$) in MeV plotted
as functions of the momentum, $k$ at densities, $\rho_0$ and 5$\rho_0$ 
for different values of the isospin asymmetry  parameter, $\eta$.
}
\end{figure}

\begin{figure}
\includegraphics[width=16cm,height=16cm]{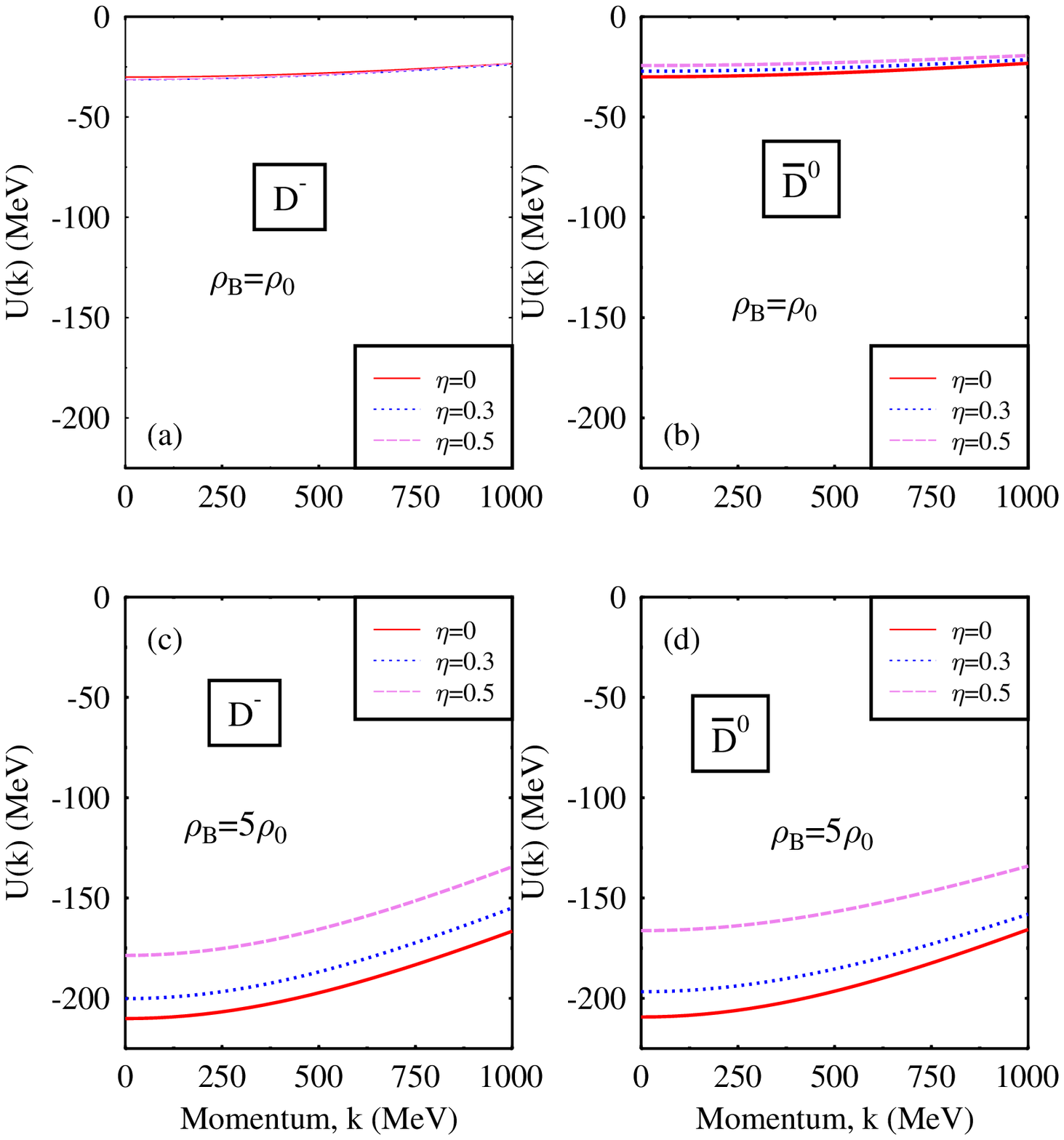}
\caption{(Color online)
The optical potentials (for $D^-$ and $\bar {D^0}$) in MeV plotted
as functions of the momentum, $k$ at densities, $\rho_0$ and 5$\rho_0$ 
for different values of the isospin asymmetry  parameter, $\eta$.
}
\end{figure}

\begin{figure}
\includegraphics[width=16cm,height=16cm]{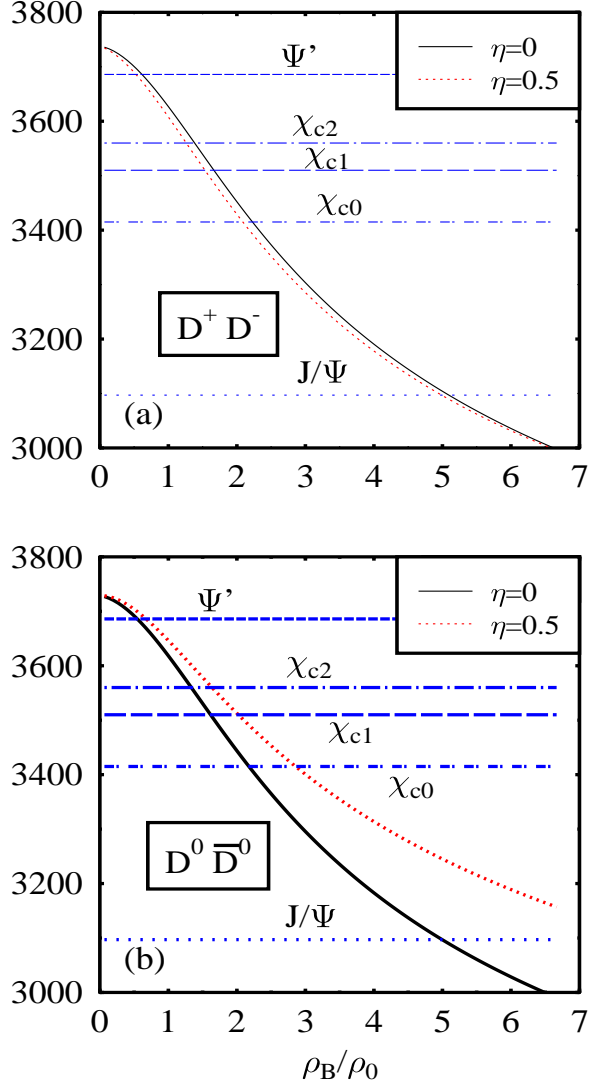}
\caption{(Color online)
The mass of the $D\bar D$ pair ($D^+D^-$ in (a) and $D^0\bar {D^0}$ in (b))
in MeV plotted as a function of $\rho_B/\rho_0$ for
values of the isospin asymmetry  parameter, $\eta$ as 0 amd 0.5.
}
\end{figure}

\end{document}